# Challenges for Real-Time Toxicity Detection in Online Games

**Authors:** Lynnette Hui Xian Ng, Adrian Xuan Wei Lim, Michael Miller Yoder

Affiliation for all authors: Collaboratory Against Hate: Research and Action Center at Carnegie Mellon University and the University of Pittsburgh

**Summary:**

Online multiplayer games like League of Legends, Counter Strike, and Skribbl.io create experiences through community interactions. Providing players with the ability to interact with each other through multiple modes also opens a Pandora's box. Toxic behaviour and malicious players can ruin the experience, reduce the player base and potentially harming the success of the game and the studio.

This article will give a brief overview of the challenges faced in toxic content detection in terms of text, audio and image processing problems, and behavioural toxicity. It also discusses the current practices in company-directed and user-directed content detection and discuss the values and limitations of automated content detection in the age of artificial intelligence.

(113 words)

**Extended abstract:**

Online multiplayer games transport players into another world and build experiences through community interactions among the players. While providing players interaction ability builds camaraderie (Perry et al., 2018), it also opens the game experience to toxic behaviour that can ruin the gameplay experience, reduce player base, potentially harming the success of the game and its studio. A survey showed that players do experience online toxicity of the forms of harassment (42%), hate speech (50%) and extremism (32%) (Unity, 2021). Therefore, toxic content moderation is essential to protect the player experience. The core of toxic content moderation is toxic content detection. The game environment hosts a series of complexity from its multiple communication modes (i.e., text, voice, images), and player behavioural signals. This talk elaborates on the challenges of toxic content detection in games, discuss the current practices, and put forth ideas for optimising toxicity detection.

Content detection in real-time requires a fine balance between automated and human detection. A team of humans have limited bandwidth, and thus increases the latency between the time the content is put forth and the time messages are validated. Thus, many companies use machine-learning AI algorithms to aid in the detection process (Fortuna & Nunes, 2018; Yousefi & Emmanouilidou, 2021), with results showing a decrease in in-game toxicity (Iain, 2020). These supervised learning algorithms need to be trained using past data, but building detection algorithms have evolved from a data and algorithm constraint to a policy constraint. With increasing online gaming, game companies and researchers can extract trove of data that can be used for training AI algorithms (e.g. chat data from League of Legends and Starcraft (Neto et al., 2017; Thompson et al., 2017)), however, some of these data have become illegal to store, e.g. the EU's GDPR enforcement on personally identifiable data for ensure the Right to be Forgotten, the Right to Consent and the Right to Explanation impacts the development of data-intensive algorithms (Humerick, 2017). Real-time content detection also means there is a significant compute cost to analyse each frame and message passed. To reduce the load on the central server, the algorithms are sometimes pushed to the player's devices, but devices like low-end mobile phones might not have enough power to process content moderation.

Text is the foundation element for building content detection algorithms because many games offer an in-game chat function (Neto et al., 2017). Current research results in machine learning algorithms like random forests focusing on racism, sexism, prejudice, hate speech etc (Fortuna & Nunes, 2018). The challenges include evolving gamer lexicon and misspellings that will evade detection (MacAvaney et al., 2019), especially so in a fast-paced environment like a multiplayer team shooter game.

Images need to be processed in games like Skribbl.io or Drawful, where users draw content as communicative devices to other players. Image content analysis involves aligning the visual image feature with textual analysis to understand the context of the drawing, and possible signals that the drawing represents (Gomez et al., 2020), keeping in mind that not all users are artistic. This differs from digital image analysis, because the freehand drawings can be fluid or deformed rather than structure, thus a higher amount of interpretation is required. It also requires analysing the images from multiple perspectives, inferring whether other users can potentially perceive a toxic image given the change in camera angles (Monroe, 2001).

Audio is culture and region sensitive. Most algorithms involve transcribing the audio then performing content moderation on the text transcripts (Padmanabhan & Johnson Premkumar, 2015). While there has been a long string of research on voice-to-text for the English languages, this field is not as studied for other languages (e.g. cantonese). With gaming slang, it is also possible to be toxic without using profanity (Sood et al., 2012), and audio filters must continually evolve with the ever-changing online slang (Reid et al., 2022).

Lastly, there is behavioural toxicity within the game. Examples are intentional feeding where the player deliberately gets killed by the opponent team, negative attitude where players deliberately disrupt the game with the intention to lose or give up and surrender, and leaving the game/AFK if they perceive the match as lost which affects the opponent's chance of winning (Kou, 2020). Currently, League of Legends defines seven behavioural toxicity for players to report. Our preliminary results surveyed attendees attending the gaming sections of the 2023 Tekko convention indicates that one of the more common form of behavioural toxicity within the game is ganging up on or intentionally excluding a player. Detection of behavioural toxicity is difficult to automate because it relies on understanding the gameplay and players' interactions with each other, and varies from game-to-game.

Some companies use third-party hate speech detection tools, yet the cost of engaging them is extremely high. For example, League of Legends uses ToxMod, an audio detection solution. At a cost of 0.10USD/hour, engaging ToxMod can be unaffordable for smaller game companies. Therefore, real-time game content detection needs to optimise a multi-modal and cost-effective solution to create a safe game environment.

Current practices in toxicity detection has two main threads: company-directed and user-directed moderation. Company-directed detection relies on a set of features that game studios put forth, i.e., Riot Games announced in May 2023 an upgrade in their text evaluation machine learning models to reduce disruptive text, and automating muting feature that prevents toxic chat messages from being sent in-game (Timtammonster, 2023). User-directed content detection allows users to report disruptive behavior for moderators to evaluate the soundness of the report and take action (e.g., limiting, banning). This technique is used in League of Legends, Counter Strike and Fortnite.

In dealing with these challenges, a multi-pronged approach is required. The cost challenge can be overcome with open-sourced tools developed with academia or non-profits. Data storage and privacy challenges could result in opt-in programs for game studios to legally record interactions, stating the use and storage information. On the machine learning side, multimodal toxicity detection systems can be developed to detect toxicity across text/audio/image/game streams. Game companies, policymakers and academia can work together to improve toxicity detection. Game studios can also focus their resources on the highly watched and streamed games, and increase the need for proactive live toxicity detection.